\documentclass[letterpaper,conference]{IEEEtran}

\usepackage{amsmath,amssymb,amsfonts}
\usepackage{algorithmic}
\usepackage{graphicx}
\usepackage{subcaption}
\usepackage{textcomp}
\usepackage{xcolor}
\usepackage{listings}
\usepackage[font=footnotesize]{caption}

\usepackage[T1]{fontenc}
\usepackage[scaled]{beramono}

\usepackage{fancyhdr}
\fancypagestyle{IEEE_Green_open_access_footer}{
  \fancyhf{}
  \fancyhead[C]{\footnotesize \copyright2019 IEEE. Personal use of this material is permitted. Permission from IEEE must be obtained for all other uses, in any current or future media, including reprinting/republishing this material for advertising or promotional purposes, creating new collective works, for resale or redistribution to servers or lists, or reuse of any copyrighted component of this work in other works.     DOI: 10.1109/LCN44214.2019.8990822}     %% C or L or R.
}

\begin{document}

\title{LIoTS: League of IoT Sovereignties.\\A Scalable approach for a Transparent Privacy-safe Federation of Secured IoT Platforms}

\author{
\IEEEauthorblockN{{\bf Flavio Cirillo}\IEEEauthorrefmark{1}\IEEEauthorrefmark{2}\\ \tt\footnotesize flavio.cirillo@neclab.eu} \and
\IEEEauthorblockN{{\bf Nicola Capuano}\IEEEauthorrefmark{2}\\ \tt\footnotesize nicol.capuano@studenti.unina.it} \and
\IEEEauthorblockN{{\bf Simon Pietro Romano}\IEEEauthorrefmark{2}\\ \tt\footnotesize spromano@unina.it} \and
\IEEEauthorblockN{{\bf Ern\"o Kovacs}\IEEEauthorrefmark{1}\\ \tt\footnotesize ernoe.kovacs@neclab.eu} \and
\and
%\IEEEauthorblockA{\hspace{150 pt}\IEEEauthorrefmark{1}NEC Laboratories Europe, Heidelberg, Germany}
%\IEEEauthorblockA{\hspace{150 pt}\IEEEauthorrefmark{2}University of Napoli Federico II, Napoli, Italy}}
\IEEEauthorblockA{
\hspace{55 pt}\IEEEauthorrefmark{1}NEC Laboratories Europe, Heidelberg, Germany;
\IEEEauthorrefmark{2}University of Napoli Federico II, Napoli, Italy}}

\maketitle
\thispagestyle{IEEE_Green_open_access_footer}

\begin{abstract}
Internet-of-Things has entered all the fields where data are produced and processed, resulting in a plethora of IoT platforms, typically cloud-based, centralizing data and services management. This has brought to many disjoint IoT silos. Significant efforts have been devoted to integration, recurrently resulting into bigger centralized infrastructures. Such an approach often stumbles upon the reluctance of IoT system owners to loose the dominion over data. We introduce a secured and privacy-safe infrastructure where a federation overlay is distributed among parties and the data control is kept locally. This establishes a league of peers each sovereign of their IoT system and data: League of IoT Sovereignties (LIoTS). LIoTS is scalable by design, allowing iterative formation of domains levels due to the transparency of its federation.
Tests show that the overhead is minimal when exchanged data is hefty, and that LIoTS performs better in large IoT deployments than centralized approaches.
\end{abstract}

\hyphenation{LIoTS}


\graphicspath{{./images/}}

\newcommand{\fc}[1]{{\color{blue}{[FC] #1}}}
\newcommand{\tbd}[1]{{\color{red}{[TBD:] #1}}}

\section{Introduction}

%\fc{ 1) a broad generally accepted statement}

The Internet-of-Things (IoT) paradigm has lately gained more and more momentum in all the fields where data are produced and processed (e.g., health care, smart cities, industry) justifying the emergence of a plethora of IoT platforms. A common approach for IoT systems deployment is to leverage the scalability and performance of a cloud-based infrastructure for storing and analyzing data~\cite{RAY2018291}. However, this brought to an abundance of single-scoped, disjoint systems generally defined as ``vertical IoT silos''.  
%\fc{2) introduce the topic, justify your study}
A lot of efforts have been devoted to define standards and ontologies to enable, on the one hand, interoperability among platforms~\cite{lanza2016fiesta,OpenIoT} and, on the other hand, inter-domain interaction~\cite{CityHub}.
%\cite{HyperCat_DataHub}.
%\fc{to check the previus citations}.
However, the harmonization of IoT systems often results in a further centralization towards another cloud instantiation~\cite{FIESTAIEEEAccess}, which leads to architecture scalability issues when handling billions of devices. In addition, IoT systems owners are reluctant to loose control over the generated data~\cite{PWC}.

%\fc{3) create your niche}

%\fc{4) state the purpose of the main activity}

This paper presents the League of IoT Sovereignties (LIoTS), a distributed infrastructure that federates IoT systems while leaving the data dominion in the hands of data owners by offering means for security and privacy control. LIoTS enables data and services brokerage of both data queries and streams among peers of IoT platforms, overcoming the fragmentation of IoT silos.
The main contribution of LIoTS is:
\begin{itemize}[]
    \item \textit{Sovereignty of data providers}: IoT providers keep data locally, thus maintaining their power over the owned data.
    \item \textit{Privacy preserved}: The privacy of intra-domain users (e.g., applications, providers, persons) is prevented to be leaked externally in the federation. This is achieved via multiple levels of security systems (intra-domain identities and policies are kept only within the domain), as well as through the IoT Registrar that automatically makes data available based on privacy directives given by the IoT providers.
    \item \textit{Scalable federation by design}: LIoTS enables the transparent existence of multiple levels of federations. This is achieved through the usage of a brokering layer for each of the federation levels, and with multiple levels of security systems.
%    \item \textit{High level of abstraction for IoT exchange}: A powerful domain-specific language for IoT data exchange is inherited by the usage of the NGSI protocol. 
    \item \textit{Plug-and-Play approach}: IoT providers and legacy systems are relieved from the burden of maintaining the federation, such as declaring or updating data availability.
\end{itemize}

\noindent Experimentation demonstrates a slight loss in terms of latency for traversing the federated and secured layer, but shows even better performance compared to a centralized approach in large scale scenarios with thousands of things, that is comparable to real IoT deployments such as SmartSantander~\cite{SANCHEZ2014217}.

\section{Background}
\label{sec:background}

LIoTS is centered around the concept of \textit{IoT context}, that is the representation of real world `status'. A context refers to an entity representing a thing (e.g., a car, a building) together with its status. The context can be physically measured by a sensor, or derived by analytics functions. 

\textbf{Federation.}
Enabling the flow of IoT contexts among platforms and IoT services in a way that is transparent to IoT actors is the key to a global Internet-of-Things overlay. A centralized approach where everything (viz. data and services) is handled by a remote authority is often not a solution for multiple reasons, such as real-time constraints, and waste of bandwidth for transmitting not requested data. 
A first approach is to have methods for the discovery of services that provide IoT context~\cite{ServiceDiscoverySurvey,OpenIoT}. Discovery is a pivotal element for seamless interoperability among systems, since not only datasets but also generic IoT services are exposed. This also enables an IoT marketplace, which is a gap identified in~\cite{MINERAUD2016IoT}. Though, discovery alone is not enough to automatize communications dispatching to the right actors. This last feature is the trait of a broker component that hides the discovery process and intermediates data flows~\cite{FogFlow}. 
This empowers a transparent federation progressing towards the intent-based programming in the IoT~\cite{PatRICIA} field.
Nevertheless, the simplicity of cloud-based IoT system is not neglected in this work, hence a hybrid approach is blended for a fully-fledged IoT infrastructure.

\textbf{Privacy and Security.}
One of the main disincentives to sharing data is the fear of loosing control over the owned information~\cite{PWC}. Thus, there is a clear need to include efficient and reliable privacy and security mechanisms. The study conducted in~\cite{DistributedIoT} assesses that most of the cloud-based IoT platforms often address typical web and network security attacks (such as DoS, eavesdropping), whereas~\cite{MINERAUD2016IoT} depicts privacy and data access control as open challenges. In this paper we address those points by offering means for IoT owners to smooth the federation of their systems while preserving privacy.

\vspace{-5pt}
\section{System Design}
\label{sec:design}
LIoTS uses disparate components playing different roles and divided in: \textit{context layer}, \textit{brokering layer} and \textit{security layer}. 

\begin{figure}
\begin{subfigure}{.25\textwidth}
  \centering
  % trim={<left> <lower> <right> <upper>}
  \includegraphics[width=1.4\columnwidth,bb= 0in 0in 10in 7.5in,trim=0cm 5cm 12cm 6.5cm,clip]{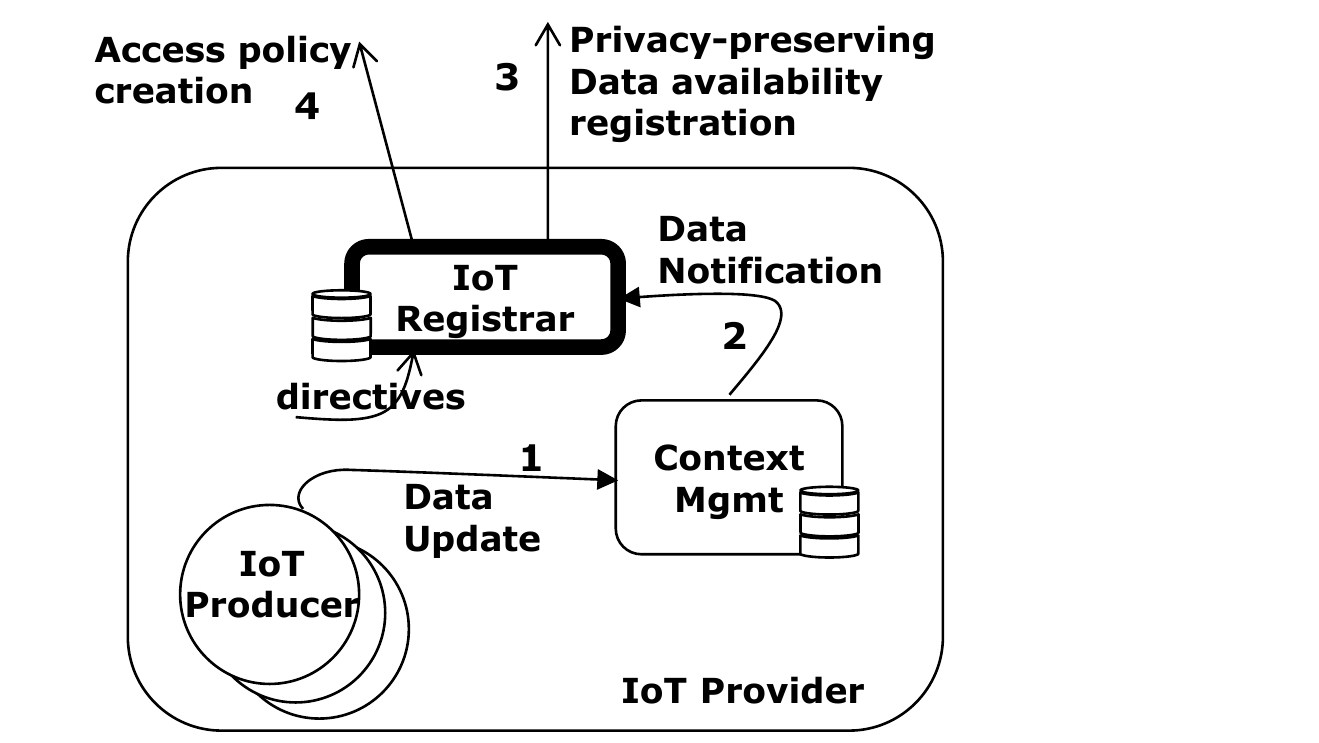}
\end{subfigure}%
%\hspace{2em}
\begin{subfigure}{.23\textwidth}
  \centering
  \includegraphics[width=1\columnwidth,bb= 0in 0in 10in 7.5in,trim=13cm 5cm 4cm 7cm,clip]{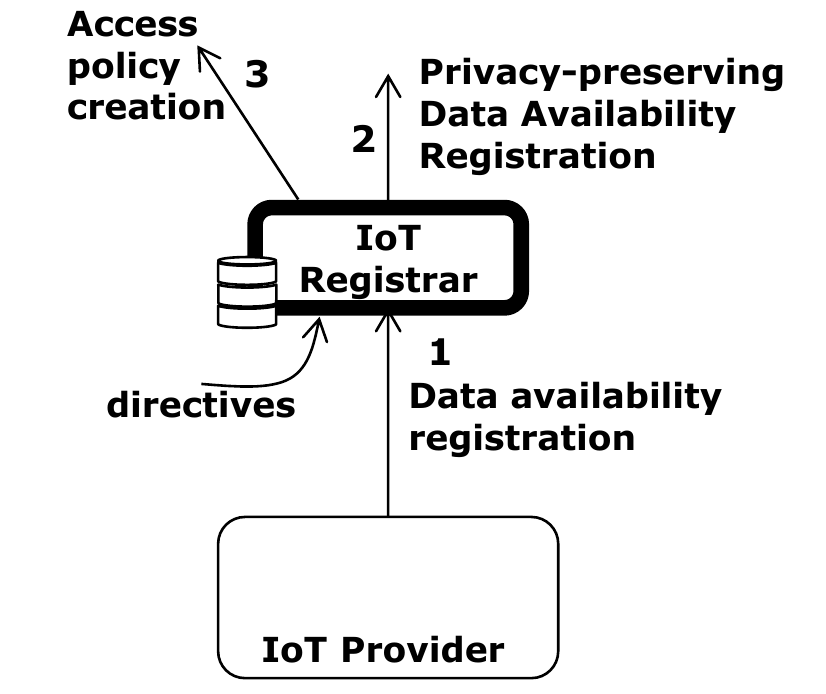}
\end{subfigure}\\
  \footnotesize{{\color{white}.}\hspace{70pt}a) \hspace{100pt}b)}
  \vspace{-5pt}
\caption{Privacy-preserving IoT Registrar: a) based on context, b) based on registrations.}
  %\vspace{-5pt}
\label{fig:iotregistrar}
%\vspace{-1em}
\end{figure}

In the context layer the \textit{IoT Producer} produces data either periodically or event-based (e.g., an environment change). Typical examples of IoT producers are sensors. 
A \textit{Context Manager (CM)} is capable of storing and indexing data and offers an interface for queries and subscriptions. An \textit{IoT Provider} is a system composed of one or more IoT producers and a CM exposing the collected data (Fig.~\ref{fig:iotregistrar}). 
Within the brokering layer, a \textit{Broker} is a mediator that, given a context request, dispatches it, transparently, to one or more IoT providers, depending on the requested context. The broker exposes methods for executing brokered queries and subscriptions. A broker needs to be assisted by a \textit{Discovery} service acting as a registry of available IoT providers together with their data. The discovery supports subscriptions and queries for IoT providers offering the requested IoT contexts.
In the security layer, the \textit{Identity Manager (IdM)} handles identities within the system. Each component in the system has an identity within the IdM, and the latter is offering a token-based protocol which allows components to identify each other. A \textit{Policy Decision Point (PDP)} responds to requests with a decision based on policies.
A \textit{Policy Enforcement Point (PEP)} secures a component by intercepting communications and imposing access policies. PEP is assisted by the IdM for authenticating requestors and %by the PDP for deciding upon policies.

\begin{figure*}
  \centering
  % trim={<left> <lower> <right> <upper>}
  \includegraphics[width=0.95\textwidth,bb= 0in 0in 13.78in 7.5in,trim=0cm 3.2cm 0cm 5.5cm,clip]{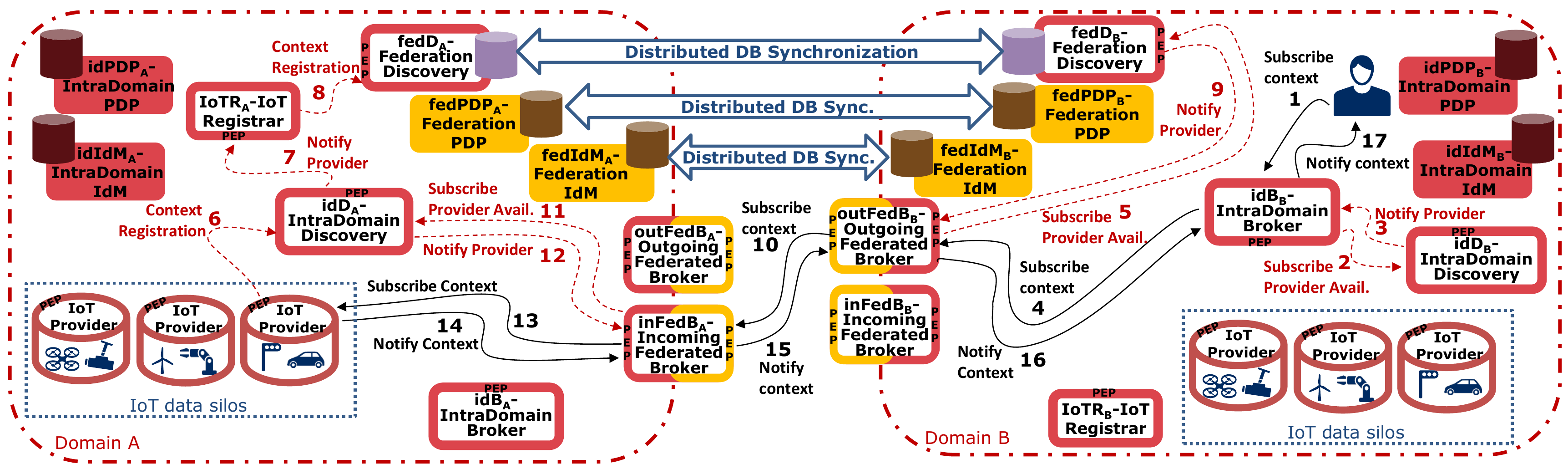}
  \vspace{-3pt}\caption{General architecture of a multi-party secured exchange platform with subscribe. Messages with no data are omitted}
  \label{fig:generalarchitecture}
  \vspace{-1.5em}
\end{figure*} 

The \textit{IoT Registrar} acts as a sort of glue between the described layer and components. It issues availability registrations to the discovery component on behalf of human administrators (see Fig.~\ref{fig:iotregistrar}). 
A registration can be generic (raising unnecessary traffic to providers) or very detailed (yet disclosing sensitive information). If such registrations are made by humans, performing detailed yet privacy-preserving registrations might require a lot of expertise for each deployment, while also inducing latency on an effective usage of the sources. Here is why a plug-and-play approach would be much more advisable. IoTR solves these problems by subscribing for information and automatically synthesizing registrations (or updating stale ones) when necessary. The IoTR bases the correct privacy preserving aggregation of information on given directives. For example it could be instructed to register sensors only by their type (hence if a second sensor of the same type gets deployed, no changes are made to the registrations set) and by loose geographical area (for example, the location of a registration refers to the nearest municipality name rather than to the exact coordinates). 
The registrar addresses also the challenge of creating access policies that is necessary for a correct behaviour of a secured IoT ecosystem~\cite{SecurityPrivacyIoT}.

%\vspace{-15pt}
\subsection{Message Flows}
\label{sec:messageflows}

The IoT data traffic typically follows $4$ paradigms: \textit{Publish-Query}, \textit{Publish-Subscribe}, \textit{Distributed Query}, \textit{Publish-Notify}. In the first paradigm an IoT producer publishes context updates to a CM which stores them. When the latter is queried it responds with the requested contexts.
In the publish-subscribe, an application first declares its interest in context data to the CM. Then context producers publish context updates and, when matching with the subscription, the CM notifies the subscriber.
In the distributed query paradigm data is not residing in a single place but needs to be procured on the fly.
When the broker receives a request, it discovers providers through the discovery, and contacts them. Then, the broker aggregates responses from providers (either into a set or by applying a function, such as averaging homologous contexts), and forwards the result to the requestor.
In different scenarios, context might be generated by a service, be measured by a sensor triggered by the request, or come from storage. 
The subscribe-notify paradigm starts with a context subscription to the broker, which subscribes to the discovery for context availability. The broker is notified about IoT providers, either in case of stored matching registrations, or as soon as a new matching one is completed. Consequently, the broker subscribes to the IoT provider(s), and collects notifications that are eventually forwarded to the context requestor.

Typically, centralized IoT platforms implement one or both the first two paradigms, implying that data is continuously flowing to the cloud regardless of the actual interest of users and/or applications. The distributed query and subscribe-notify paradigms, instead, foresee ``lazy'' data flows as the data is pulled from remote providers on demand. In order to keep the simplicity of the centralized paradigm also for the distributed paradigms, we make usage of the IoT Registrar (Fig.~\ref{fig:iotregistrar}).

\subsection{Multi-party exchange platform: system architecture}
\label{sec:architecture}

The system architecture of a secured multi-party IoT data exchange platform is shown in Fig.~\ref{fig:generalarchitecture} with an illustrative federated publish-notify data exchange. 
Each provider silo manages its own IoT deployment and handles data within its premises or its cloud of choice, exposing them through a CM (see Fig.~\ref{fig:iotregistrar}a).
One or more IoT providers are clustered in domains. Every domain has two independent security systems for two scopes of action: \textit{intra-domain} and \textit{federation}.
The intra-domain security system, coloured in red in Fig.~\ref{fig:generalarchitecture}, is formed by an IdM (idIdM), a PDP (idPDP), and a PEP for each component to be secured against intra-domain access. When a message arrives at a secured component carrying an access token, the PEP first checks the authenticity of the sender through the idIdM and then their access rights with the idPDP. If any of the checks fails, access is denied; otherwise, the message is forwarded to the secured component. The federation security system is formed by a federation IdM (fedIdM) and PDP (fedPDP) for each domain, as well as a PEP for every component exposed externally. The fedIdMs (as fedPDPs) have synchronized databases such that the same request to any of them results in the same response. 
How the synchronization is achieved (e.g., off-the-shelf distributed databases such as Cassandra, blockchain-based technologies) is out of the scope of this paper. 
Every domain possesses an identity for the authentication and authorization between domains. The federation access policies are visible to all parties.
At setup time, components need to make configurations (Tab.~\ref{tab:federationsettings}) for enabling the federated communication.

For intra-domain data exchange the behaviours and message flows are similar to the one presented in \S\ref{sec:messageflows}, with the brokerage of an intra-domain Broker (idB) assisted by an intra-domain Discovery (idD) service. When the data requested is available in a different domain, other components enter the game. 
Every domain has a federation Discovery (fedD) with the database synchronized. Thus, registrations stored in a fedD are visible to anybody in the federation, and, therefore, it is of utmost importance to control the content of registrations. Each domain has an IoT Registrar (IoTR) that accepts registrations from providers and synthesizes new ones for the fedD (Fig.~\ref{fig:iotregistrar}b).
For brokering messages between domains each domain deploys two broker instances: the incoming federation Broker (inFedB) handles requests coming from outside the domain, while the outgoing federation Broker (outFedB) looks after requests going in the outward direction. 
The two federation brokers are associated with two different discoveries. When a request comes from outside the domain the inFedB discovers data providers within the domain against the idD (see Tab.\ref{tab:federationsettings}). When a request comes from inside the domain, instead, the outFedB needs to contact the fedD for discovering other domains providing the data of interest. Therefore all the registrations done by the IoTR to the fedD must carry the exposed address of the inFedB (or its PEP). Each of these boundary brokers are protected by two different PEPs, one assisted by the intra-domain security system with the aim of regulating who in the domain may do federated requests, the other assisted by the federation security system for moderating requests from external parties. Having two disjoint security layers allows domain administrators to decide which section of data can be exposed. In other words, a PEP of an IoT provider might treat a request coming from the inFedB differently than a request from within the domain. 
In addition, if domain\textsubscript{B} has different federation access policies among domain\textsubscript{B} users, the outFedB\textsubscript{B} receives unfiltered data, but the PEP\textsubscript{idB\textsubscript{B}} refines data as prescribed in idPDP\textsubscript{B}. This approach permits to hide a user\textsubscript{B} existence and associated policies outside of the domain. 
Thus, separating the security systems has the benefit to protect sensitive information contained inside policies, as well as identities pertaining only to the domain. 

\begin{table}[htbp]
\begin{center}
\setlength\tabcolsep{2pt}
\vspace{-3pt}
\caption{LIoTS infrastructure settings: \textit{d} for domain, \textit{fed} for federation.}
\vspace{-5pt}
\label{tab:federationsettings}
\begin{tabular}{|c|c|c|c|c|c|c|}
\hline
                                                                   & \textbf{idB} & \textbf{idD} & \textbf{IoTR}                                                    & \textbf{outFedB}                                                                              & \textbf{inFedB}                                                                            & \textbf{fedD} \\ \hline\hline

\textbf{\begin{tabular}[c]{@{}c@{}}Security\\ System\end{tabular}} & \textit{d}       & \textit{d}       & \textit{d}                                                           & \begin{tabular}[c]{@{}c@{}}\textit{d} for query\\ and subscribe;\\ \textit{fed} for notify\end{tabular} & \begin{tabular}[c]{@{}c@{}}\textit{d} for notify;\\ \textit{fed} for query\\ and subscribe\end{tabular} & \textit{d}        \\ \hline
\textbf{\begin{tabular}[c]{@{}c@{}}\textbf{Registra-}\\\textbf{tion}\end{tabular}}                                              & none         & none         & none                                                             & \begin{tabular}[c]{@{}c@{}}as provider\\for  everything\\to idD\end{tabular}                   & \begin{tabular}[c]{@{}c@{}}as provider for\\all registrations\\ in fedD\end{tabular}    & none          \\ \hline
\textbf{\begin{tabular}[c]{@{}c@{}}\textbf{Subscrip-}\\\textbf{tion}\end{tabular}}                                              & none         & none         & \begin{tabular}[c]{@{}c@{}}provider\\avail. to idD\end{tabular} & none                                                                                          & none                                                                                       & none          \\ \hline
\textbf{Discovery}                                                 & idD          & n/a          & n/a                                                              & fedD                                                                                          & idD                                                                                        & n/a           \\ \hline
\end{tabular}
\end{center}
\vspace{-10pt}
\end{table}

\begin{figure}
  \centering
  % trim={<left> <lower> <right> <upper>}
  \includegraphics[width=0.95\columnwidth,bb= 0in 0in 11.81in 9.84in,trim=0cm 6cm 0cm 1.9cm,clip]{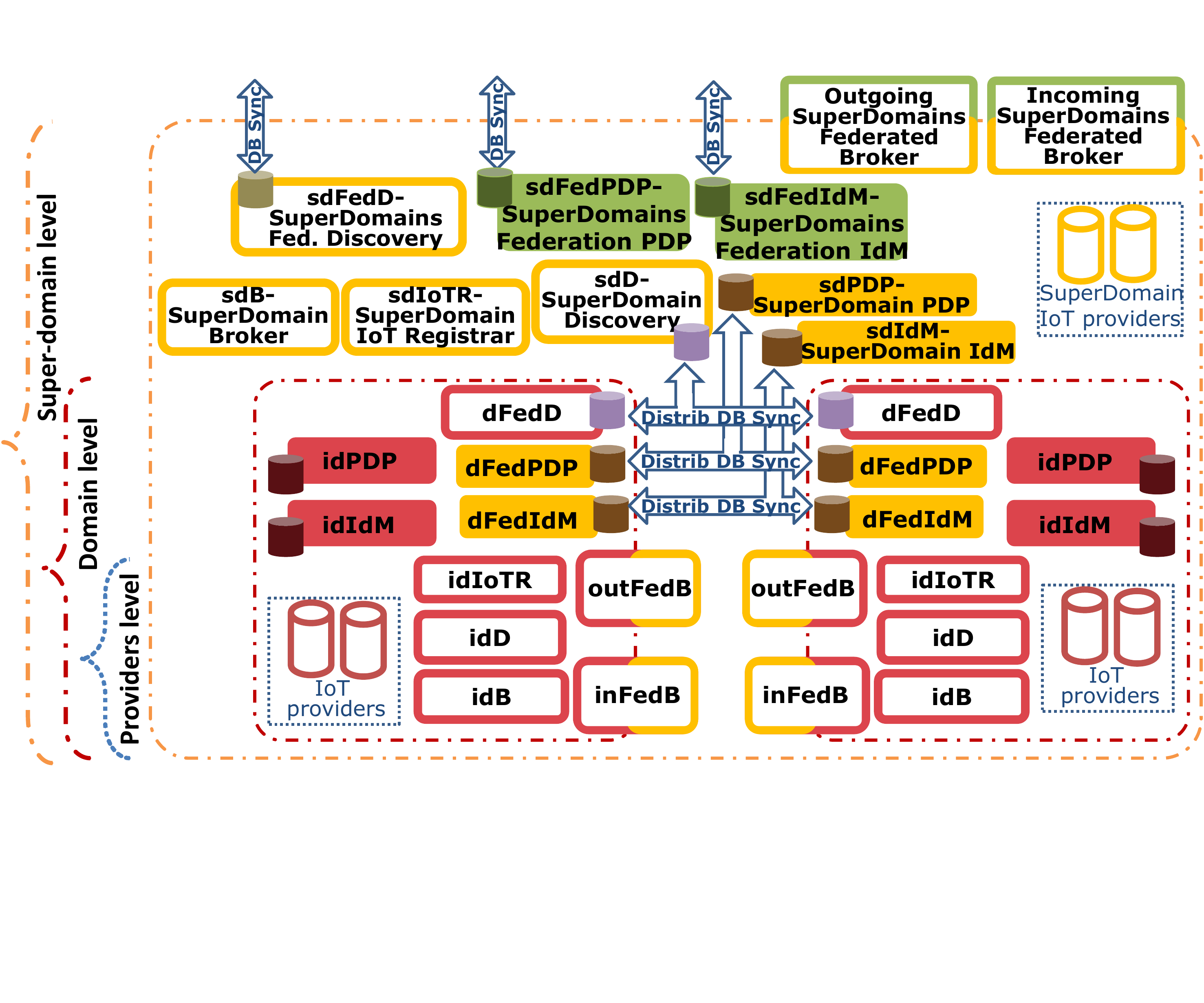}
   %\vspace{-5pt}
   \caption{ Scaling the proposed architecture iteratively.}
   \label{fig:scaledArch}
%\vspace{-5pt}
\end{figure}

The proposed architecture can be scaled for achieving super-domains of domains iteratively. For this purpose, each domain of Fig.~\ref{fig:generalarchitecture} is seen as a domain IoT provider by the super-domain infrastructure (see Fig.~\ref{fig:scaledArch}). 
In this case there are three separated security layers: the intra-domain, super-domain, and super-domains federation. LIoTS is easily adaptable to different scenarios and configurations. It is possible to have hybrid solutions where IoT providers are directly managed by the super-domain. In Fig.~\ref{fig:scaledArch} there are $3$ levels of federation
but, theoretically, the number of levels is unbounded.

\vspace{-3pt}
\section{Evaluation}
\label{sec:evaluation}

\begin{figure}
  \centering
  \includegraphics[width=0.95\columnwidth]{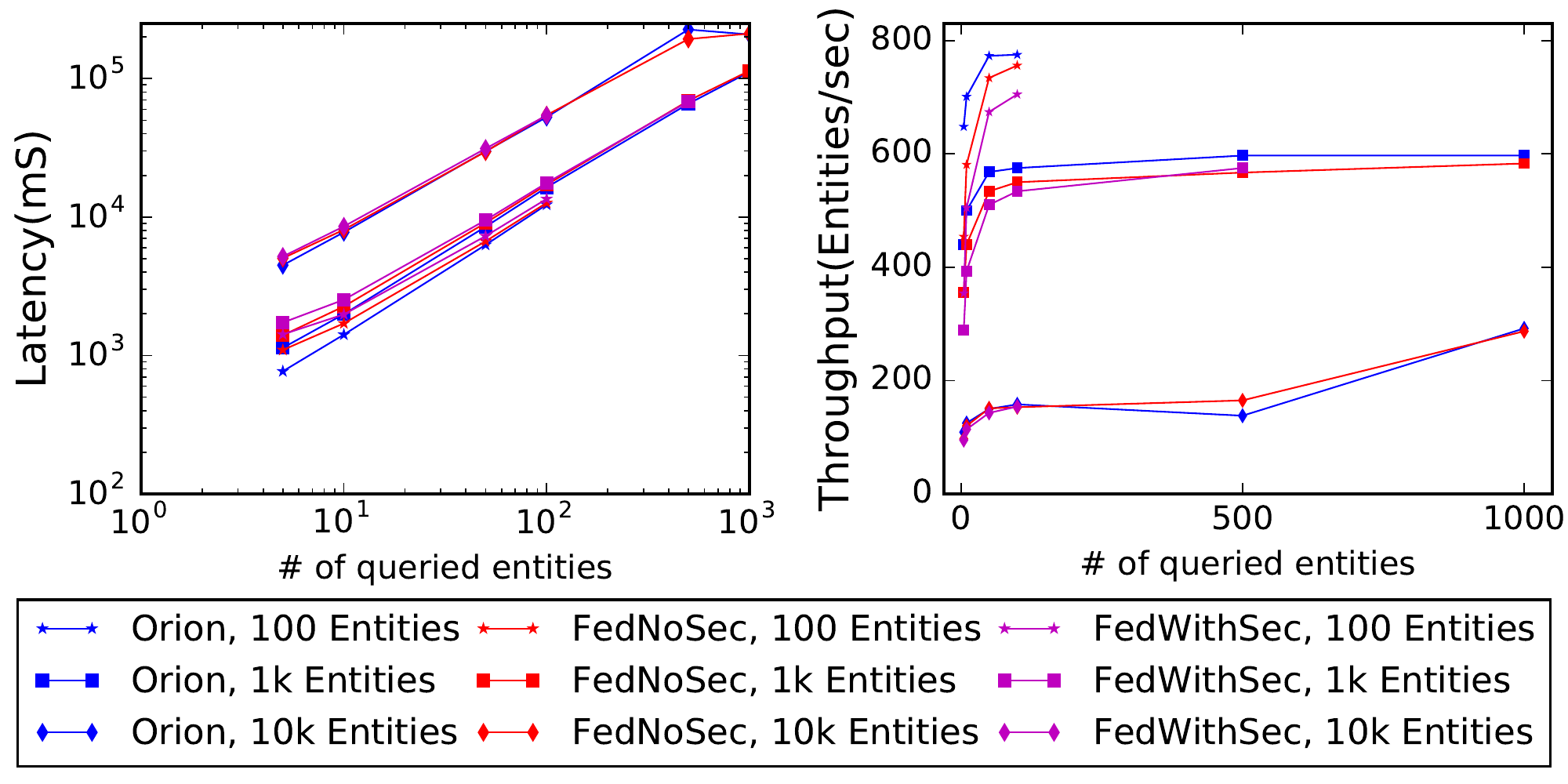}
  \vspace{-5pt}
\caption{Latency and throughput for query scenarios}
  \label{fig:query}
  \vspace{-5pt}
\end{figure}

We tested the distributed query paradigm on both secured and unsecured (i.e., lacking the security layer) federated architectures, in all cases against the unsecured publish-query paradigm representing a centralized architecture. 
The envisioned scenario is that applications within domain\textsubscript{B} request data residing in domain\textsubscript{A}. 
As testing components we have used, from the FIWARE framework, Orion as CM, IoT Broker as Broker, NEConfMan as Discovery, AuthZForce as PDP, KeyRock as IdM, and Wilma as PEP. In addition, we implemented a prototype of IoT Registrar.
Apache JMeter is used to perform query requests for a random number of randomly chosen entities. The test is carried out varying the number of total entities handled by the CM (100, 1000 and 10000), representing the size of the IoT deployment if we consider each entity as a `thing'. The number of attributes per stored entity and the number of attributes queried per entity are fixed to 100 and 20, respectively.
We have taken 10000 as the IoT deployment top size taking into consideration that SmartSantander~\cite{SANCHEZ2014217} handles 20000 entities.
The throughput is normalized by considering the amount of information represented by entity contexts returned in each query response, e.g., if in a scenario with 100 queried entities per request we achieve 20 requests/sec, normalization brings to a throughput of 2000 entities/sec. For fair comparison, test points with a non-zero error rate are omitted from the graphs. 
In Fig.~\ref{fig:query} the different colour lines for both latency and throughput are getting closer as the number of queried entities increases, meaning that the overhead becomes more and more negligible. This is even more noticeable as the dimension of the deployment increases, hence indicating the CM as the system's bottleneck. 

\begin{figure}
  \centering
  \vspace{-5pt}
  \includegraphics[width=0.95\columnwidth]{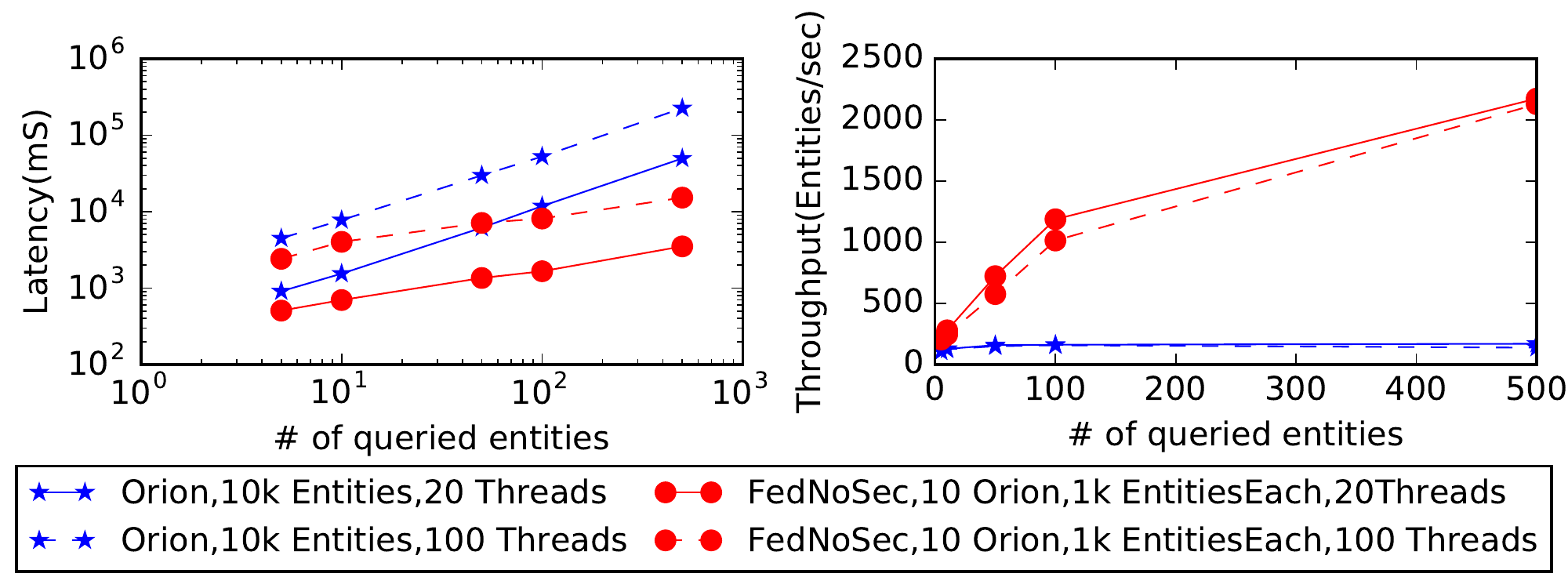}
  \vspace{-5pt}
  \caption{Results for multi- and single-provider scenarios}
  \label{fig:multiorion}
\end{figure}

We performed one more test to investigate the benefit of load balancing implicit in a distributed architecture. We compared the federated architectures (unsecured and secured) comprising 10 CMs, each handling 1000 entities, with a single unsecured centralized CM with 10000 entities. We vary also the number of concurrent requesting clients (threads) between 20 and 100. Each of the small CM handles disjoint sets of entities whilst the big CM handles them all. We have then performed randomized queries and therefore for each query 1 or more CM need to be contacted by the Broker.
Fig.\ref{fig:multiorion} shows that a federated architecture performs much better than a centralized approach, in terms of both throughput and latency.

%\vspace{-5pt}
\section{Conclusions and Future Works}
\label{sec:conclusions}

In this paper we have presented a distributed federation architecture that is scalable by design. Privacy and security play a big role in the overall system. The evaluation shows that overhead introduced in big scale deployments 
is negligible and the federation approach is even better performing in multi-provider scenarios. Each of the providers is considered as a different organization holding data locally and protecting their sovereignty over owned information.

Enforcing obligations on the data usage is the next step for a secured IoT data exchange and we plan to start from LIoTS to design a distributed data usage control system.

\bibliographystyle{IEEEtran}
{\footnotesize
\bibliography{Bib}
}

\end{document}